\documentclass[epj,twocolumn]{webofc}
\usepackage[varg]{txfonts}   
\woctitle{Hadron Collider Physics symposium 2012}

\begin{document}
\title{Searches for MSSM Higgs Bosons at ATLAS and CMS}
%
%

\author{Stan Lai\inst{1}\fnsep\thanks{\email{stan.lai@cern.ch}}, on behalf of the ATLAS and CMS Collaborations
}

\institute{Albert-Ludwigs Universit\"{a}t Freiburg 
          }

\abstract{%
The Minimal Supersymmetric extension of the Standard Model (MSSM) predicts the existence of 
three neutral and two charged Higgs bosons. Searches for these MSSM Higgs bosons are presented, 
based on proton-proton collisions recorded in 2011 and 2012 by the ATLAS and CMS experiments at the LHC.
The neutral Higgs bosons are searched through their decays into pairs of oppositely charged tau leptons.  
The exclusion limits at 95\% confidence level are shown as a function of the $m_A$ and 
$\tan\beta$ parameters. The search for the charged Higgs bosons is 
based on their production through the decays of top quarks in the $t\bar{t}$ process, $t \to bH^+$. 
The Higgs bosons subsequently decay predominantly into a tau lepton and a neutrino. Upper limits are 
set on the branching fraction $B(t \to bH^+)$, combining the final states with leptonic and 
hadronic tau decay modes.

}
\maketitle
\section{Introduction}
\label{sec:intro}

Discovering the mechanism responsible for electroweak symmetry breaking
is a major goal of the physics programme at the Large Hadron Collider~(LHC)~\cite{LHC}.
In the Standard Model (SM), this mechanism requires the existence of a scalar particle:
the Higgs boson~\cite{Higgs}.  The recent discovery of a particle compatible with the 
Higgs boson at the LHC~\cite{HiggsDiscovery} provides evidence in support of this picture. 
Even if this recently discovered particle is shown to have properties entirely consistent with the 
SM Higgs boson, there are still a number of problems not yet addressed by the SM. 
For instance, quantum corrections to the mass of the Higgs boson contain quadratic divergences.
This problem can be solved by introducing supersymmetry, a 
symmetry between fermions and bosons, by which the divergent corrections to the Higgs 
boson mass are cancelled.

In the Minimal Supersymmetric Standard Model (MSSM)~\cite{MSSM},
two Higgs doublets are necessary, coupling separately to up-type and down-type fermions.
This results in five physical Higgs bosons, two of which are neutral and CP-even 
($h$, $H$), one of which is neutral and CP-odd ($A$), and two of which are charged ($H^\pm$). 
At tree level their properties can be described in terms of two parameters,
typically chosen to be the mass of the CP-odd Higgs boson, $m_A$, and the ratio of
the vacuum expectation values of the two Higgs doublets, $\tan\beta$.  
In the MSSM, the Higgs boson couplings to  $\tau$ leptons and $b$-quarks
are strongly enhanced for a large part of the parameter space.
This is especially true for large values of $\tan\beta$, in which case the
decay of a Higgs boson to a pair of $\tau$ leptons and its production
in association with $b$-quarks play a much more important role than in the SM.

\section{Charged Higgs Searches}
\label{sec:Charged}

Charged Higgs bosons are predicted in the MSSM as well as in non-minimal Higgs scenarios such as Two Higgs Doublet
Models~\cite{2HDM}, and the observation of $H^\pm$ would clearly indicate new physics beyond the SM.
In the MSSM, the main production mode for $H^\pm$ at the LHC is through top quark decays $t \to bH^+$
for $m_{H^+} < m_\mathrm{top}$~\footnote{Henceforth, charged Higgs bosons are denoted $H^+$, with the charge-conjugate $H^-$ implied.}.
The dominant source for top quarks at the LHC is $t\bar{t}$ production.  For $\tan\beta >2$,
the decay $H^+ \to \tau\nu$ is dominant.

\subsection{ATLAS search for $H^+ \to \tau\nu$}
\label{subsec:ATLAStaunu}

The ATLAS search for $H^+ \to \tau\nu$ in the mass range $90 < m_{H^+} < 160$~GeV uses $t\bar{t}$ events
with a leptonically or hadronically decaying $\tau$ lepton in the final state~\cite{ATLAStaunu}.  
The results are based on 4.6~fb$^{-1}$ of data from 7 TeV $pp$ collisions.  Three final states are analyzed:
\begin{itemize}
\item lepton+jets: $t\bar{t} \to b\bar{b}WH^+ \to b\bar{b}(q\bar{q}^\prime)(\tau_\mathrm{lep}\nu)$
\item $\tau$+lepton: $t\bar{t} \to b\bar{b}WH^+ \to b\bar{b}(\ell\nu)(\tau_\mathrm{had}\nu)$
\item $\tau$+jets: $t\bar{t} \to b\bar{b}WH^+ \to b\bar{b}(q\bar{q}^\prime)(\tau_\mathrm{had}\nu)$.
\end{itemize}

The lepton+jets search uses events passing a single lepton trigger and requires one electron with
$E_\mathrm{T} > 25$~GeV or one muon with $p_\mathrm{T} > 20$~GeV.  In addition, four jets with 
$p_\mathrm{T} > 20$~GeV are required, with exactly two of them identified as originating from $b$-quarks.
A requirement of $E_\mathrm{T}^\mathrm{miss} > 40$~GeV is applied, and additionally 
$E_\mathrm{T}^\mathrm{miss} \times |\sin\Delta\phi_{\ell, \mathrm{miss}}| > 20$ GeV is required if
the azimuthal angle $\Delta\phi_{\ell, \mathrm{miss}}$ between the lepton and $E_\mathrm{T}^\mathrm{miss}$
is less than $\pi/6$.

The hadronic side of the event is identified by selecting the combination of $b$-tagged jets ($b$) and two
untagged jets ($j$) that minimizes
\begin{equation}
\chi^2 = \frac{(m_{jjb} - m_\mathrm{top})^2}{\sigma^2_\mathrm{top}} + \frac{(m_{jj} - m_W)^2}{\sigma_W^2}, \nonumber
\end{equation}
where $\sigma_\mathrm{top} = 17$~GeV and $\sigma_W = 10$~GeV are the widths of the reconstructed top quark
and W boson mass distributions.  Events with $\chi^2 > 5$ are rejected.

Backgrounds with misidentified leptons are evaluated in a data-driven estimate, using a control sample
with looser lepton identification than the default selection.  Using known efficiencies for identifying
a real or fake lepton measured in $Z\to \ell\ell$ and multijet events respectively, the number of events
with misidentified leptons can be estimated.

Finally, the following requirements are imposed:
\begin{equation}
\cos \theta^*_\ell = \frac{2m_{b\ell}^2}{m^2_\mathrm{top} - m^2_W} -1 < -0.6 \nonumber
\end{equation}
and
\begin{equation}
m_\mathrm{T}^W = \sqrt{2p^\ell_\mathrm{T}E^\mathrm{miss}_\mathrm{T}(1 - \cos\Delta\phi_{\ell, \mathrm{miss}})}. \nonumber
\end{equation}
The variable $m_\mathrm{T}^H$ defined by
\begin{equation}
(m_\mathrm{T}^H)^2 = \left(\sqrt{m^2_\mathrm{top} + (\vec{p}_\mathrm{T}^\ell + \vec{p}_\mathrm{T}^b + \vec{p}_\mathrm{T}^\mathrm{miss} )^2} 
                        - p_\mathrm{T}^b \right)^2 - \left( \vec{p}_\mathrm{T}^\ell +  \vec{p}_\mathrm{T}^\mathrm{miss}  \right)^2 \nonumber
\end{equation}
is used as the final discriminating variable in this channel.

The $\tau$+lepton search also uses events passing a single lepton trigger and requires one electron with
$E_\mathrm{T} > 25$~GeV or one muon with $p_\mathrm{T} > 20$~GeV.  A hadronically decaying $\tau$ candidate
($p_\mathrm{T} > 20$~GeV) with opposite charge to the lepton is also required.  At least two jets 
($p_\mathrm{T} > 20$~GeV) are required, at least one of which is $b$-tagged.  Finally, the sum of all 
track transverse momenta $\sum p_\mathrm{T}^\mathrm{tracks}$ for tracks ($p_\mathrm{T} > 1$~GeV) 
associated with the primary vertex is required to be greater than 100 GeV.

Backgrounds with fake leptons are evaluated in the same way as the lepton+jets channel.  Backgrounds with 
electrons and jets misidentified as $\tau_\mathrm{had}$ candidates are evaluated also in a data-driven estimate.  The 
misidentification probability for electrons as $\tau_\mathrm{had}$ candidates is measured using a $Z\to ee$ control 
sample, and are applied to all simulated events in the analysis.  A control sample enriched in $W$+jets 
events is used to measure the probability for jets to be misidentified as $\tau_\mathrm{had}$ candidates, and these 
probabilities are then applied on simulated background samples.  The differences in jet composition 
between $t\bar{t}$ and $W$+jets events on this misidentification probability is taken into account 
as a systematic uncertainty.

As a final discriminating variable, the $E_\mathrm{T}^\mathrm{miss}$ variable is used for the limit 
setting, to be discussed later.  


The $\tau$+jets search uses events passing a $\tau + E_\mathrm{T}^\mathrm{miss}$ trigger.  The event 
selection requires at least four jets ($p_\mathrm{T} > 20$~GeV), of which at least one is $b$-tagged, 
and one hadronically decaying $\tau$ candidate with $p_\mathrm{T} > 40$~GeV.  In addition, 
$E_\mathrm{T}^\mathrm{miss} > 65$~GeV is required, and events must also satisfy
\begin{equation}
\frac{E_\mathrm{T}^\mathrm{miss}}{0.5 \mathrm{GeV}^{1/2} \sqrt{\sum p_\mathrm{T}^\mathrm{tracks} }} > 13. \nonumber
\end{equation}
The combination of the $b$-tagged jet ($b$) and two untagged jets ($j$) with the highest 
$p_\mathrm{T}^{jjb}$ must satisfy $120 < m_{jjb} < 240$~GeV, to pick out topologies consistent 
with top quark decays.

Backgrounds with electrons and jets misidentified as $\tau_\mathrm{had}$ candidates are estimated
using the same method as the $\tau$+lepton search.
In addition, multijet backgrounds are estimated by fitting the $E_\mathrm{T}^\mathrm{miss}$ distribution in data, where the
shape of the distribution for multijet events is determined in a control sample where the $\tau$ candidates fail the
tight identification requirements of the default selection, and no $b$-tagged jets are found.  For the evaluation of 
backgrounds with real hadronically decaying $\tau$ leptons, the embedding method is used~\cite{ATLASembed}.  
Here, a sample of $t\bar{t}$ $\mu$+jets events are selected, and the detector signature of the muon is 
replaced by a simulated hadronic $\tau$ decay, providing a data-driven estimate for such backgrounds.

\begin{figure}
\centering

\includegraphics[width=6.5cm, clip]{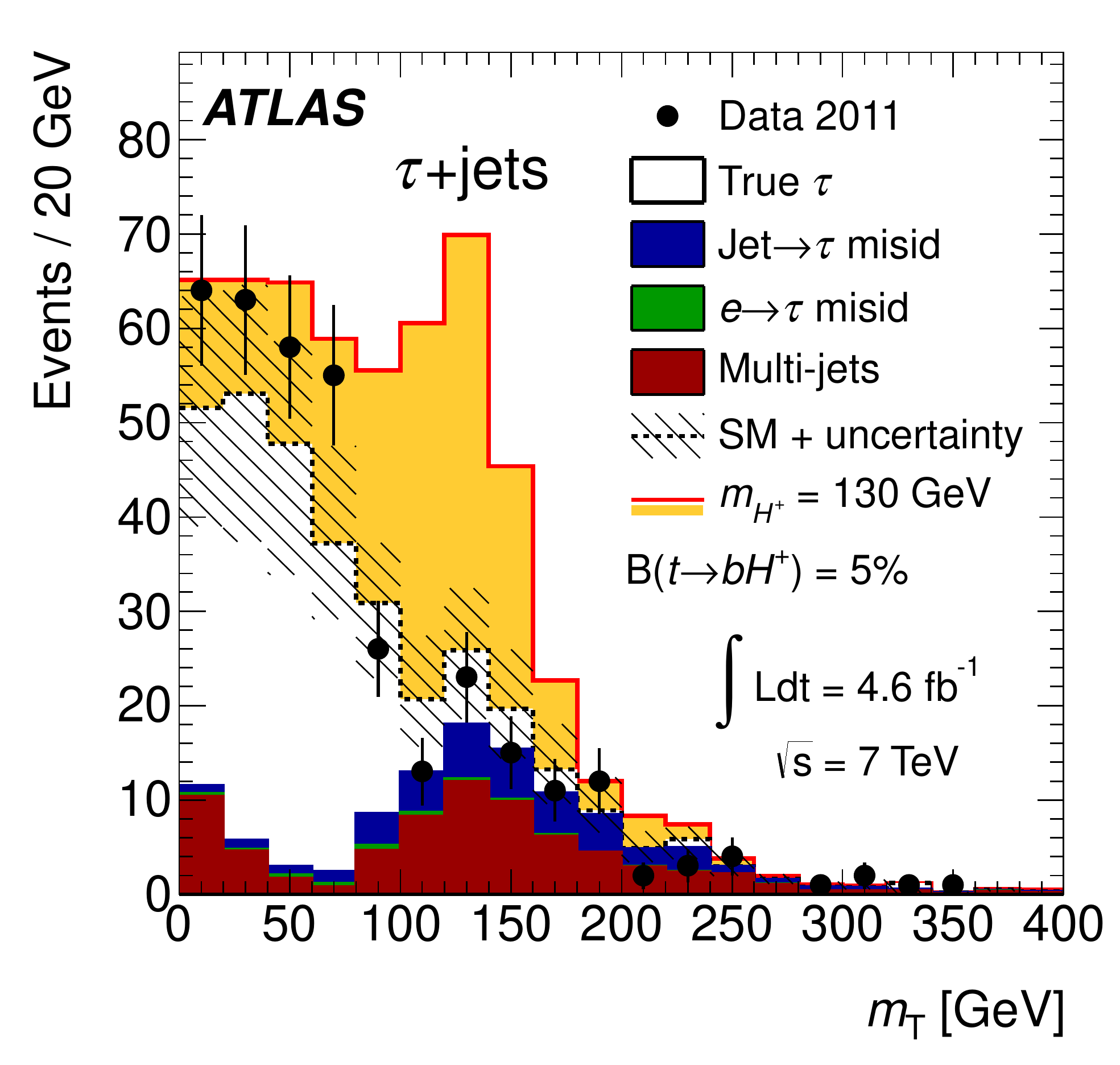}
\caption{The $m_\mathrm{T}$ distribution in the $\tau$+jets final state.  The dashed line corresponds 
to the SM-only hypothesis.  The solid line shows the predicted contribution
of signal+background for a 130 GeV charged Higgs boson with $B(t \to bH^+) = 5\%$ and 
$B(H^+ \to \tau\nu) = 100\%$~\cite{ATLAStaunu}.}
\label{fig:charged2}       
\end{figure}

As a final discriminating variable, the transverse mass 
$m_\mathrm{T} = \sqrt{2 p_\mathrm{T}^\tau E_\mathrm{T}^\mathrm{miss} (1 - \cos\Delta\phi_{\tau,\mathrm{miss}})}$ of the
$\tau_\mathrm{had}$ candidate and the $E_\mathrm{T}^\mathrm{miss}$ is used.  
The distribution of this variable in the selected data, along with the predicted background and signal ($m_{H^+} = 130$~GeV) 
distributions are shown in Figure~\ref{fig:charged2}.

\begin{figure*}
\centering
\includegraphics[width=6.5cm, clip]{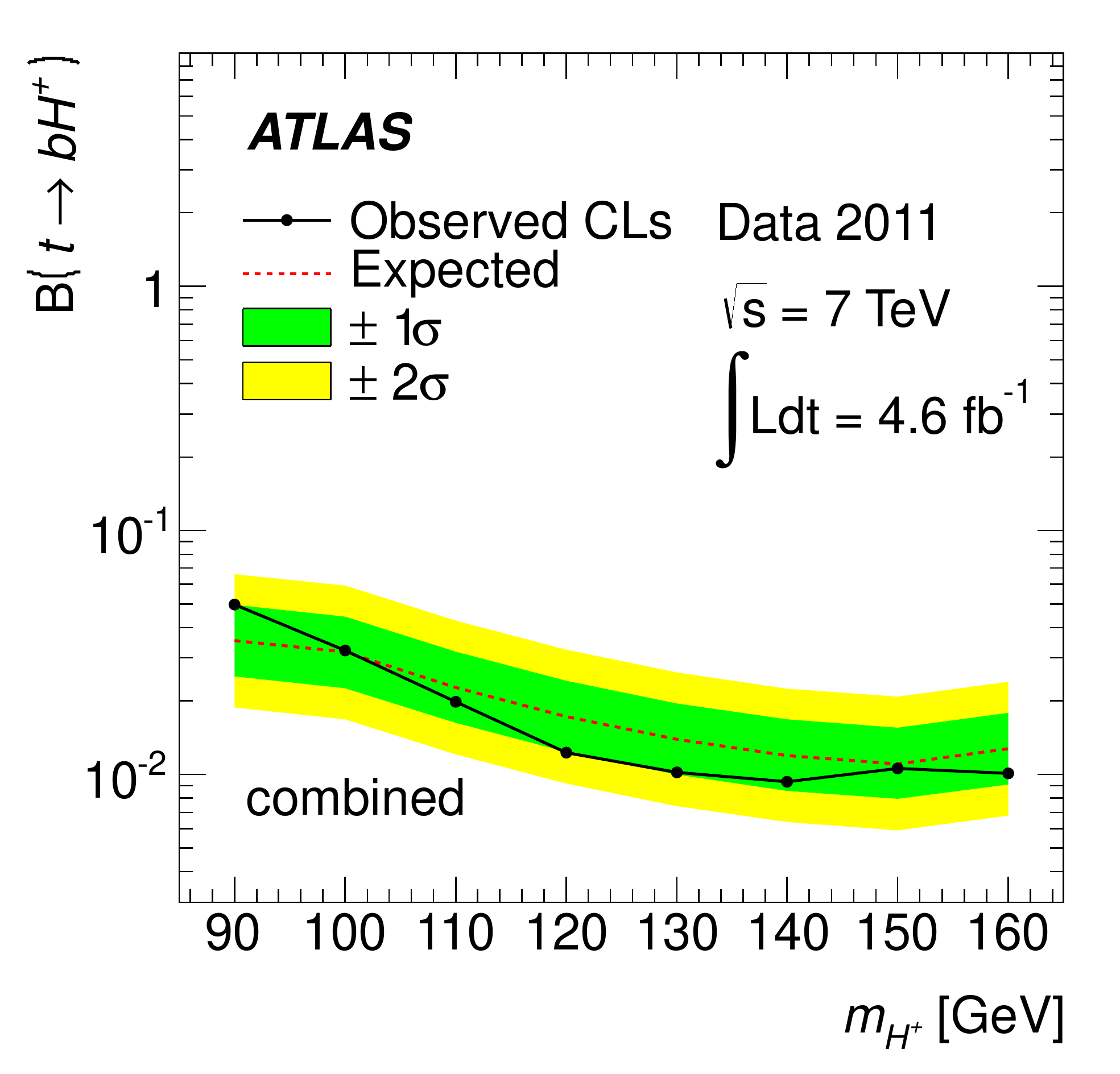}
\includegraphics[width=6.5cm, clip]{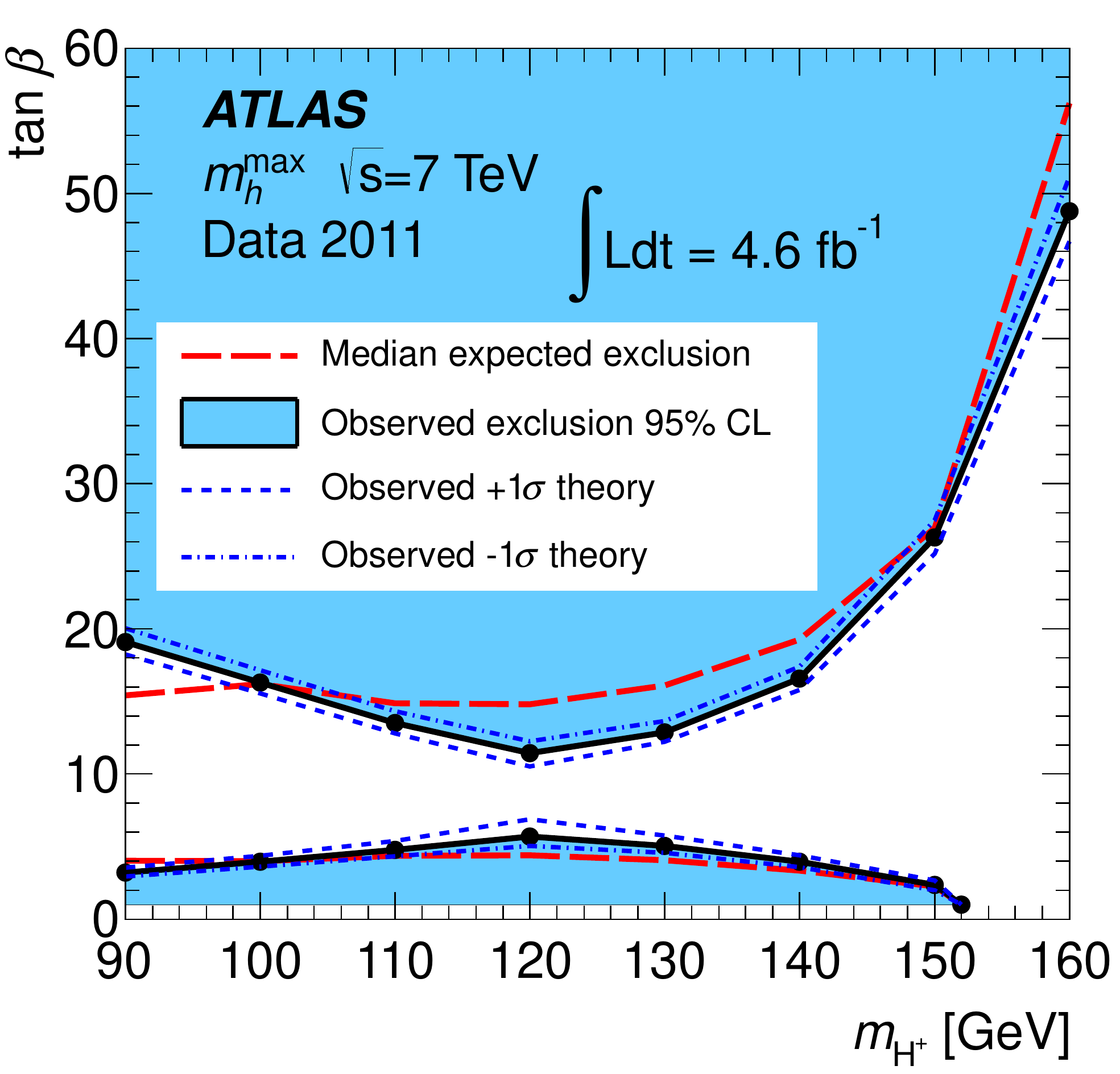}

\caption{Left: ATLAS expected and observed 95\% CL exclusion limits on $B(t \to bH^+)$ as a function of $m_{H^+}$, 
               assuming $B(H^+ \to \tau\nu) = 100\%$.
         Right: 95\% CL exclusion limits on $\tan\beta$ as a function of $m_{H^+}$, shown for the MSSM $m_h^\mathrm{max}$
                scenario.~\cite{ATLAStaunu}}
\label{fig:charged3}       
\end{figure*}

A profile likelihood ratio is used on the final discriminating variables on all three channels to test the compatibility
of the data with the background-only and signal+background scenarios.  As no significant deviation from the SM prediction
is seen, exclusion limits at 95\% confidence level (CL) are set on the branching ratio $B(t \to bH^+)$ and in the 
$m_{H^+}$ vs $\tan\beta$ plane for the $m_h^\mathrm{max}$ scenario~\cite{mhmax}.  These are shown in Figure~\ref{fig:charged3}.

\subsection{CMS search for $H^+ \to \tau\nu$}
\label{subsec:CMStaunu}

The CMS search for  $H^+ \to \tau\nu$ in the mass range $80 < m_{H^+} < 160$~GeV also uses $t\bar{t}$ events
with a leptonically or hadronically decaying $\tau$ lepton~\cite{CMStaunu}.  Results are based on $2.0\mbox{--}2.3$~fb$^{-1}$
of 7 TeV collision data.  Three final states are analyzed:
\begin{itemize}
\item $e\mu$: $t\bar{t} \to b\bar{b}WH^+ \to b\bar{b}(\ell^\prime \nu)(\tau_\mathrm{lep}\nu)$
\item $\tau$+lepton: $t\bar{t} \to b\bar{b}WH^+ \to b\bar{b}(\ell\nu)(\tau_\mathrm{had}\nu)$
\item $\tau$+jets: $t\bar{t} \to b\bar{b}WH^+ \to b\bar{b}(q\bar{q}^\prime)(\tau_\mathrm{had}\nu)$.
\end{itemize}

The $e\mu$ search uses events passing an $e\mu$ trigger, and at least one isolated electron and 
at least one isolated muon with $p_\mathrm{T} > 20$~GeV are required.  The invariant mass of the oppositely charged 
electron-muon pair must also satisfy $m_{e\mu} > 12$~GeV.  In addition, 
two jets with $p_\mathrm{T} > 30$~GeV are required.
After all selection criteria have been applied, the dominant background comes from $t\bar{t}$ dilepton
events, followed by $Z/\gamma^* \to \ell\ell$ and single top production.

The $\tau$+lepton search selects events firing a single muon trigger, or an electron+jets trigger.  
Events are selected by requiring one isolated electron (muon) satisfying $p_\mathrm{T} > 35(30)$~GeV.  
A hadronically decaying $\tau$ candidate ($p_\mathrm{T} > 20$~GeV) and at least two jets are required, 
at least one of which is $b$-tagged.  In addition a requirement $E_\mathrm{T}^\mathrm{miss} > 45(40)$~GeV 
is placed for the $e\tau_\mathrm{had}$ ($\mu\tau_\mathrm{had}$) final state.

Backgrounds with misidentified $\tau_\mathrm{had}$ candidates are estimated by measuring the $\tau_\mathrm{had}$
misidentification probability in a control sample of $W$+jets and multijet events.  Other backgrounds 
are estimated using simulation, and the dominant background consists of $t\bar{t}$ events with SM decay modes.

The $\tau$+jets search at CMS uses events passing a $\tau + E_\mathrm{T}^\mathrm{miss}$ trigger.  
A 1-prong $\tau_\mathrm{had}$ candidate with $p_\mathrm{T}>40$~GeV is required, along with three jets 
($p_\mathrm{T} > 30$~GeV), at least one of which is $b$-tagged.  In addition, requirements are
placed on 
$E_\mathrm{T}^\mathrm{miss} > 50$~GeV and $\Delta \phi(\tau_\mathrm{had}, E_\mathrm{T}^\mathrm{miss}) < 160^\circ$.

\begin{figure}
\centering
\includegraphics[width=6.5cm, clip]{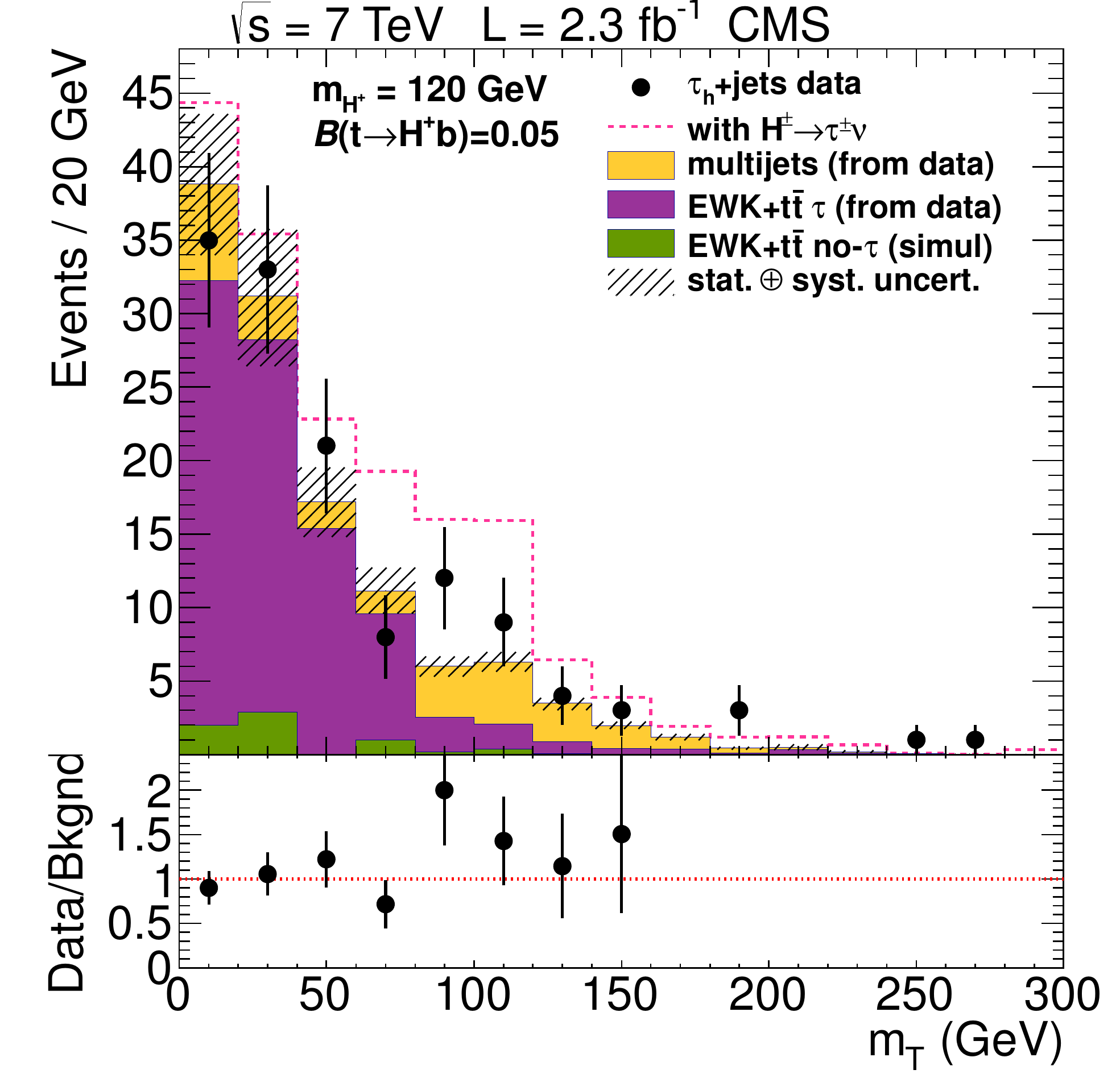}
\caption{The $m_\mathrm{T}$ distribution in the $\tau$+jets final state.  The solid line corresponds 
to the SM-only hypothesis.  The dashed line shows the predicted contribution
of signal+background for a 120 GeV charged Higgs boson with $B(t \to bH^+) = 5\%$ and $B(H^+ \to \tau\nu) = 100\%$~\cite{CMStaunu}.}
\label{fig:charged4}       
\end{figure}

This particular channel uses the transverse mass, $m_\mathrm{T}$, of the $\tau_\mathrm{had}$ and 
$E_\mathrm{T}^\mathrm{miss}$ system as a final discriminating variable.  
This distribution is shown in Figure~\ref{fig:charged4}.  
The shape and normalization of the $m_\mathrm{T}$ distribution are obtained using data-driven techniques for 
multijet and backgrounds with true $\tau_\mathrm{had}$ leptons in the final state.  For multijet background, 
this is obtained using a control sample where the requirements on $\tau_\mathrm{had}$ candidate isolation 
and the presence of a $b$-tagged jet have been lifted.  For backgrounds with a true $\tau_\mathrm{had}$ lepton 
in the final state, the embedding method is used to determine the normalization and $m_\mathrm{T}$ shapes, 
as in the ATLAS search.

\begin{figure*}
\centering

\includegraphics[width=6.5cm, clip]{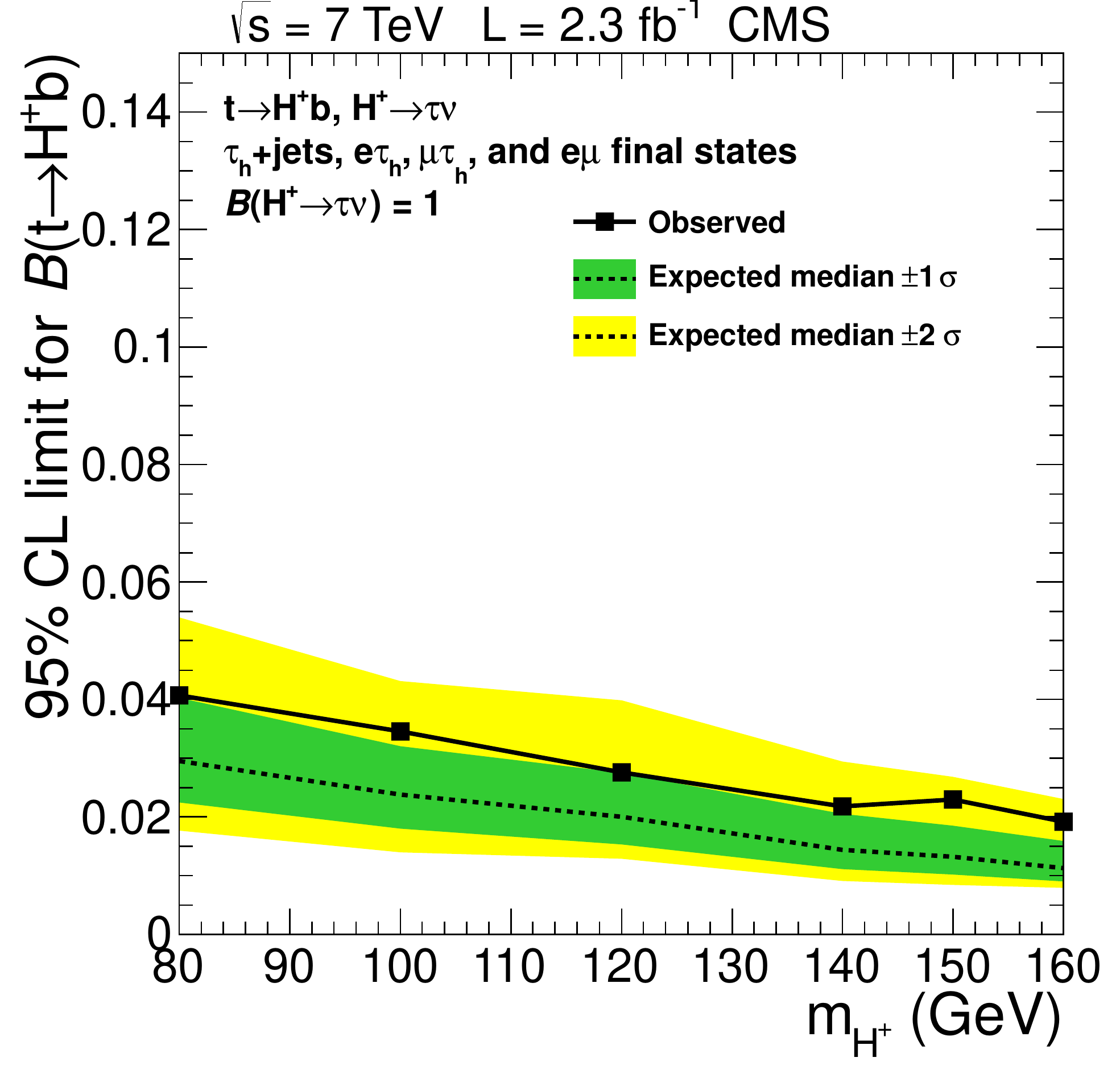}
\includegraphics[width=6.5cm, clip]{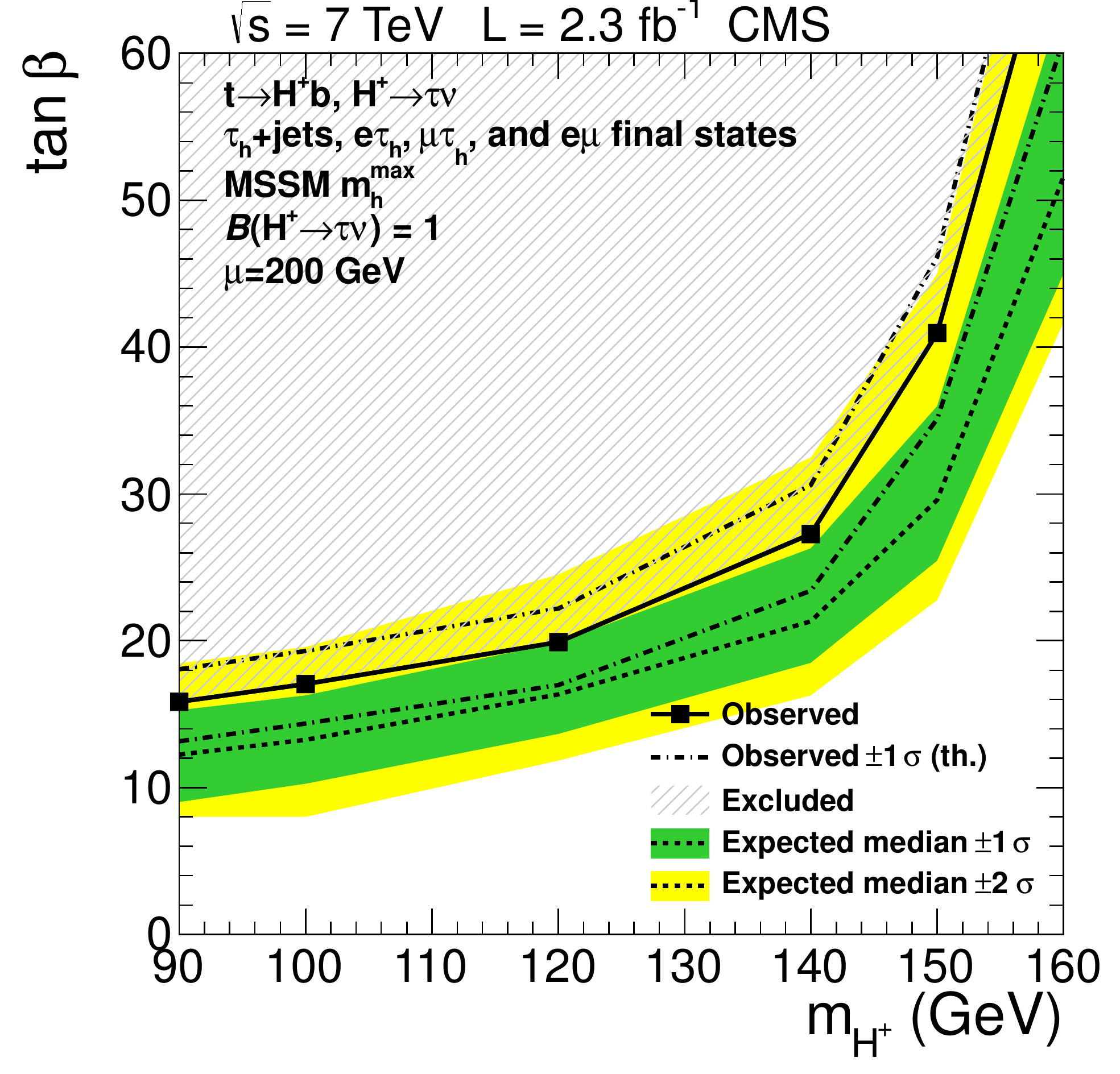}

\caption{Left: CMS expected and observed 95\% CL exclusion limits on $B(t \to bH^+)$ as a function of $m_{H^+}$, assuming $
               B(H^+ \to \tau\nu) = 100\%$.
         Right: 95\% CL exclusion limits on $\tan\beta$ as a function of $m_{H^+}$, shown for the MSSM $m_h^\mathrm{max}$
         scenario.~\cite{CMStaunu}}
\label{fig:charged5}       
\end{figure*}

Exclusion limits are extracted on $B(t \to bH^+)$ for all three channels at 95\% CL, as no significant 
excess of events over the SM prediction has been observed.  For the extraction of these limits, the $\tau$+jets 
channel uses a binned profile likelihood on the $m_\mathrm{T}$ distribution, while the other two channels 
contribute as counting experiments.  These limits as well as exclusion limits in the $m_{H^+}$ vs 
$\tan\beta$ plane for the $m_h^\mathrm{max}$ scenario are shown in Figure~\ref{fig:charged5}.

\section{Neutral Higgs Searches}
\label{sec:Neutral}

The most common production modes for neutral Higgs bosons\footnote{$\Phi$ will be used as a general symbol for the MSSM neutral Higgs bosons $h, H, A$.} 
$\Phi$ at the LHC are the $b$-quark annihilation and gluon-fusion processes.  Both processes have 
cross sections that increase with $\tan \beta$, with $b$-quark associated production becoming dominant
at high values of $\tan \beta$.  Requiring the presence of $b$-tagged jets in the final state can
thus increase the sensitivity to this search.  The most common decay modes for MSSM neutral Higgs bosons
are $\Phi \to b\bar{b}$ ($B \sim 90\%$) and $\Phi \to \tau\tau$ ($B \sim 10\%$).

Both ATLAS and CMS searches for $\Phi \to \tau\tau$ divide the analysis into subchannels based on
the decay modes of the $\tau$:
$\tau_e\tau_\mu$, $\tau_\mu\tau_\mu$ (CMS only), 
$\tau_\mathrm{lep}\tau_\mathrm{had}$, and $\tau_\mathrm{had}\tau_\mathrm{had}$ (ATLAS only).
The main background for all search subchannels is the irreducible $Z/\gamma^* \to\tau\tau$
background, which is evaluated with the use of the aforementioned data-driven embedding technique.
In the embedding technique for neutral Higgs searches, a sample of $Z \to \mu\mu$ events is selected in data, 
with the detector signature of the muons replaced by simulated hadronic $\tau$ decays.

\subsection{ATLAS search for $\Phi \to \tau\tau$}
\label{subsec:ATLAStautau}

The ATLAS search for $\Phi \to \tau\tau$ uses $4.7$~fb$^{-1}$ of 7 TeV $pp$ collision data 
accumulated in 2011~\cite{ATLAStautau}.  For the $\tau_e\tau_\mu$ subchannel, events are selected 
using either a single lepton trigger or a combined $e\mu$ trigger.  The isolated electron (muon) 
is selected with $p_\mathrm{T} > 15 (10)$~GeV, and the electron and muon candidates must be oppositely 
charged while satisfying $m_{e\mu} > 30$~GeV and transverse angle separation $\Delta\phi_{e\mu} > 2.0$.

The signal region is then divided into a $b$-tagged region (if one $b$-tagged jet with $p_\mathrm{T} > 20$~GeV
is found), and a jet-vetoed region (requiring no jets with $p_\mathrm{T} > 20$~GeV in the final state).
For the $b$-tagged region additional requirements include 
$E_\mathrm{T}^\mathrm{miss} + p_\mathrm{T}^e + p_\mathrm{T}^\mu < 125$~GeV, 
$\sum\limits_{\ell = e, \mu} \cos \Delta\phi_{E_\mathrm{T}^\mathrm{miss}, \ell} > -0.2$,
and that the scalar sum of jet transverse energies $H_\mathrm{T}$ is below $100$~GeV.

Backgrounds from multijet and $t\bar{t}$ production in the $\tau_\mathrm{e}\tau_\mathrm{\mu}$ subchannel 
are evaluated using data-driven techniques.  Backgrounds from multijet production are evaluated using
a sample with anti-isolation requirements on electron and muon candidates, while the $t\bar{t}$ background is 
evaluated in a control region with two $b$-tagged jets and no requirement on $H_\mathrm{T}$.
The final discriminating variable for the $b$-tagged region is the $\tau\tau$ invariant
mass, reconstructed with the Missing Mass Calculator (MMC) algorithm~\cite{MMC}, while the jet-vetoed
analysis defines a window in $m_{e\mu}$ to count events.  Here, the MMC algorithm accounts for
the role of the missing energy from the neutrinos in the final state by scanning the possible values
for the neutrino 4-momentum-vectors, and using the most probable value in the invariant mass
calculation.

For the $\tau_\mathrm{lep}\tau_\mathrm{had}$ search, events are selected using single lepton
triggers.  In addition to one electron or muon candidate, a hadronically decaying $\tau$ candidate
($p_\mathrm{T} > 20$~GeV) with opposite sign is required.   The transverse mass of the lepton
candidate and the $E_\mathrm{T}^\mathrm{miss}$ must satisfy $m_\mathrm{T} < 30$~GeV.  A $b$-tagged
signal region is defined if a $b$-tagged jet with $20< p_\mathrm{T} < 50$~GeV is found, and a
$b$-vetoed signal region is defined when there is no $b$-tagged jet with $p_\mathrm{T} >20$~GeV.

\begin{figure}
\centering

\includegraphics[width=6.5cm, clip]{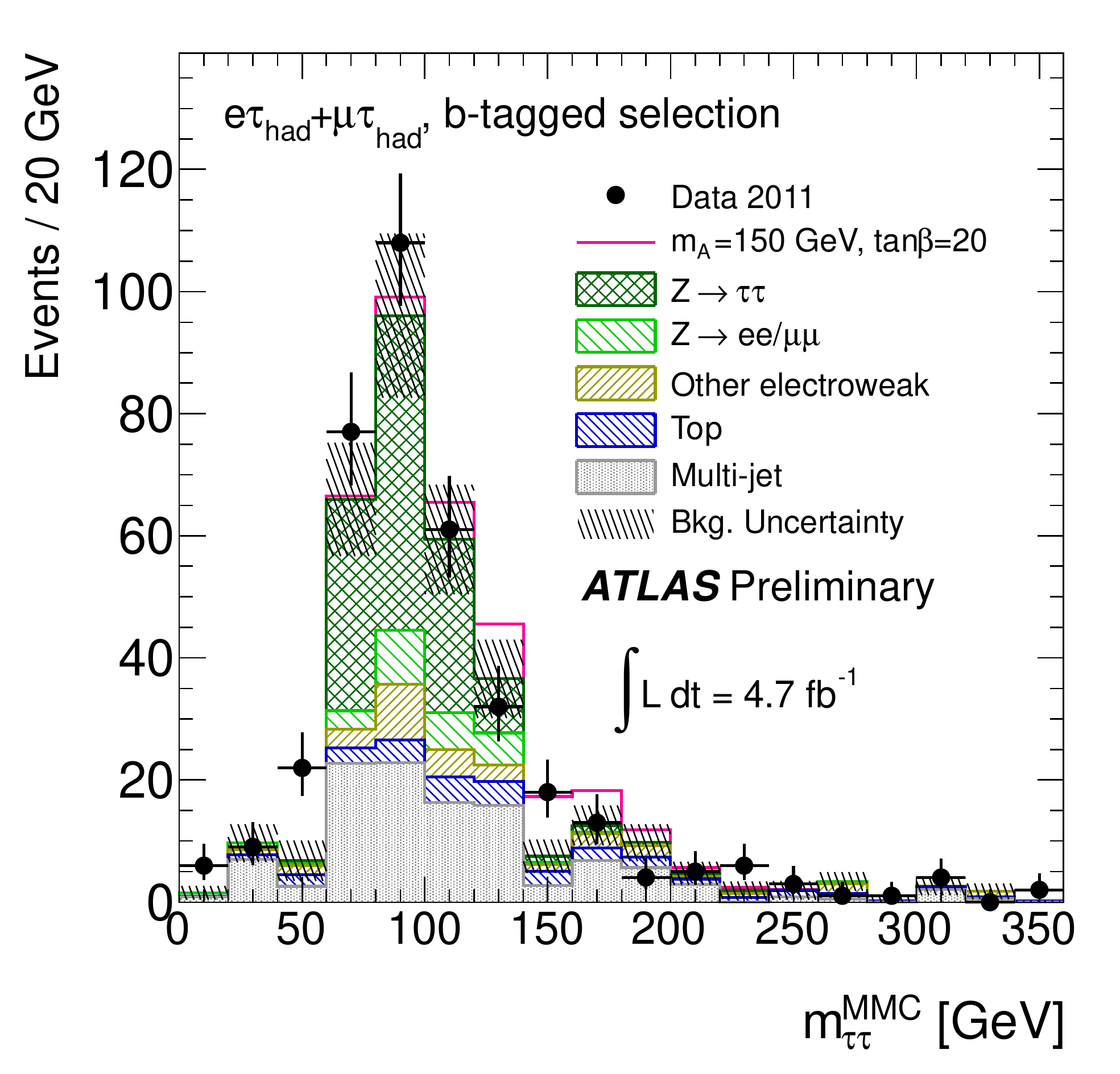}
\caption{The MMC mass shown for the $b$-tagged $\tau_\mathrm{lep}\tau_\mathrm{had}$ selection.
The data are compared to the background expectation and a hypothetical MSSM signal with $m_A = 150$~GeV
and $\tan\beta = 20$~\cite{ATLAStautau}.}
\label{fig:neutral1}       
\end{figure}

Backgrounds from $W$+jets, $t\bar{t}$, $Z/\gamma^*$ and multijet production are all evaluated using
data-driven techniques.  In particular, the probability for jets to be misidentified as $\tau_\mathrm{had}$ candidates
is measured in a region enriched in $W$+jets events ($70 < m_\mathrm{T} < 110$~GeV), and correction
factors to the simulated misidentification rate are determined and applied to $W$+jets events in the
signal region.  The $t\bar{t}$ background is normalized using a control region with two $b$-tagged jets,
where the leading $b$-tagged jet is required to have $p_\mathrm{T} > 50$~GeV.
The final discriminating variable in the $\tau_\mathrm{lep}\tau_\mathrm{had}$
channel is the MMC mass, which is shown for the $b$-tagged selection in Figure~\ref{fig:neutral1}.

The $\tau_\mathrm{had}\tau_\mathrm{had}$ search selects events that fired a ditau trigger.  Two oppositely
charged $\tau_\mathrm{had}$ candidates are required with one of the candidates passing a stricter set of $\tau$ 
identification criteria than the other.  A $b$-tagged and $b$-vetoed signal region are defined as in the 
$\tau_\mathrm{lep}\tau_\mathrm{had}$ search, and the transverse momentum thresholds for the $\tau_\mathrm{had}$ candidates
are 45~GeV (60~GeV) and 30~GeV respectively for the $b$-tagged ($b$-vetoed) selection.

The dominant background comes from multijet production, which is estimated in data using control regions
where the two $\tau_\mathrm{had}$ candidates are not oppositely charged, or where they fail the $\tau$ identification
requirements.  The MMC mass is also the final discriminating variable in this subchannel.

\begin{figure}
\centering

\includegraphics[width=6.5cm, clip]{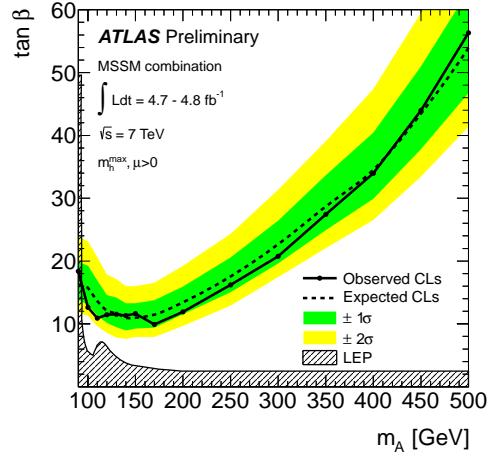}
\caption{Expected and observed 95\% confidence level exclusion limits as a function of $m_A$ and $\tan\beta$
for the $m_h^\mathrm{max}$ scenario for the ATLAS $\Phi \to \tau\tau$ search~\cite{ATLAStautau}.}
\label{fig:neutral2}       
\end{figure}

As no significant excess of events is seen in any of the subchannels, 95\% CL exclusion limits are set
on the $\Phi \to \tau\tau$ production cross section and $B(\Phi \to \tau\tau)$ by using a profile
likelihood on the MMC mass distribution (except for the $\tau_e\tau_\mu$ jet-vetoed selection).  
This is interpreted as exclusion limits as a function of $m_A$ and $\tan\beta$ in the $m_h^\mathrm{max}$ 
scenario, shown in Figure~\ref{fig:neutral2}.

\subsection{CMS search for $\Phi \to \tau\tau$}
\label{subsec:CMStautau}

The CMS search for $\Phi \to \tau\tau$ uses $4.9$~fb$^{-1}$ of 7 TeV data and $12.1$~fb$^{-1}$ of 8 TeV data 
accumulated in 2011 and 2012, respectively~\cite{CMStautau}.  All subchannels use the maximum likelihood 
mass to reconstruct $m_{\tau\tau}$~\cite{MaxLikeMass}, which is the final discriminating variable.
The maximum likelihood mass accounts for the missing energy of the neutrinos by choosing the
kinematics that maximize a likelihood function.
The $E_\mathrm{T}^\mathrm{miss}$ used for the analysis includes a multivariate regression
correction for pileup effects.
All subchannels are split into a $b$-tagged category and $b$-vetoed category, based on whether a $b$-tagged
jet ($p_\mathrm{T} >20$~GeV) is present in the final state.

For the $\tau_\mathrm{e}\tau_\mathrm{\mu}$ and $\tau_\mathrm{\mu}\tau_\mathrm{\mu}$ searches, two oppositely
charged leptons are required with thresholds of $p_\mathrm{T} > 20 (10)$~GeV for the (second) highest
$p_\mathrm{T}$ candidate.  In addition, a requirement on $p_\zeta - 0.85 p_\zeta^\mathrm{vis} > -25$~GeV 
is applied where $p_\zeta = p_\mathrm{T, \ell_1} \cdot \zeta + p_\mathrm{T, \ell_2} \cdot \zeta + p_\mathrm{T}^\mathrm{miss} \cdot \zeta$
and $p_\zeta^\mathrm{vis} = p_\mathrm{T, \ell_1} \cdot \zeta + p_\mathrm{T, \ell_2} \cdot \zeta$.  Here, $p_\mathrm{T, \ell_2}$
and $p_\mathrm{T, \ell_2}$ are the transverse momenta of the two lepton candidates, while
$\zeta$ denotes the axis that bisects the vectors parallel to these transverse momenta.
This requirement discriminates against $t\bar{t}$ and other background from SM electroweak processes.
Backgrounds with fake lepton candidates are evaluated using control regions where the lepton selection
is relaxed.

For the $\tau_\mathrm{lep}\tau_\mathrm{had}$ search, one electron or muon
is required, and an oppositely charged $\tau_\mathrm{had}$ candidate with $p_\mathrm{T} > 20$~GeV is also required.  For 
this search, the transverse mass $m_\mathrm{T}$ of the lepton and $E_\mathrm{T}^\mathrm{miss}$ is required
to be less than 40~GeV.  The normalization of backgrounds from multijet or $W$+jet events where there is a 
jet misidentified as a $\tau_\mathrm{had}$ candidate are evaluated using control samples enriched in $W$+jet events 
with large transverse mass as well as a region enriched in multijet events where the $\tau_\mathrm{had}$ candidates 
are required to have the same electric charge.

\begin{figure*}
\centering

\includegraphics[width=6.5cm, clip]{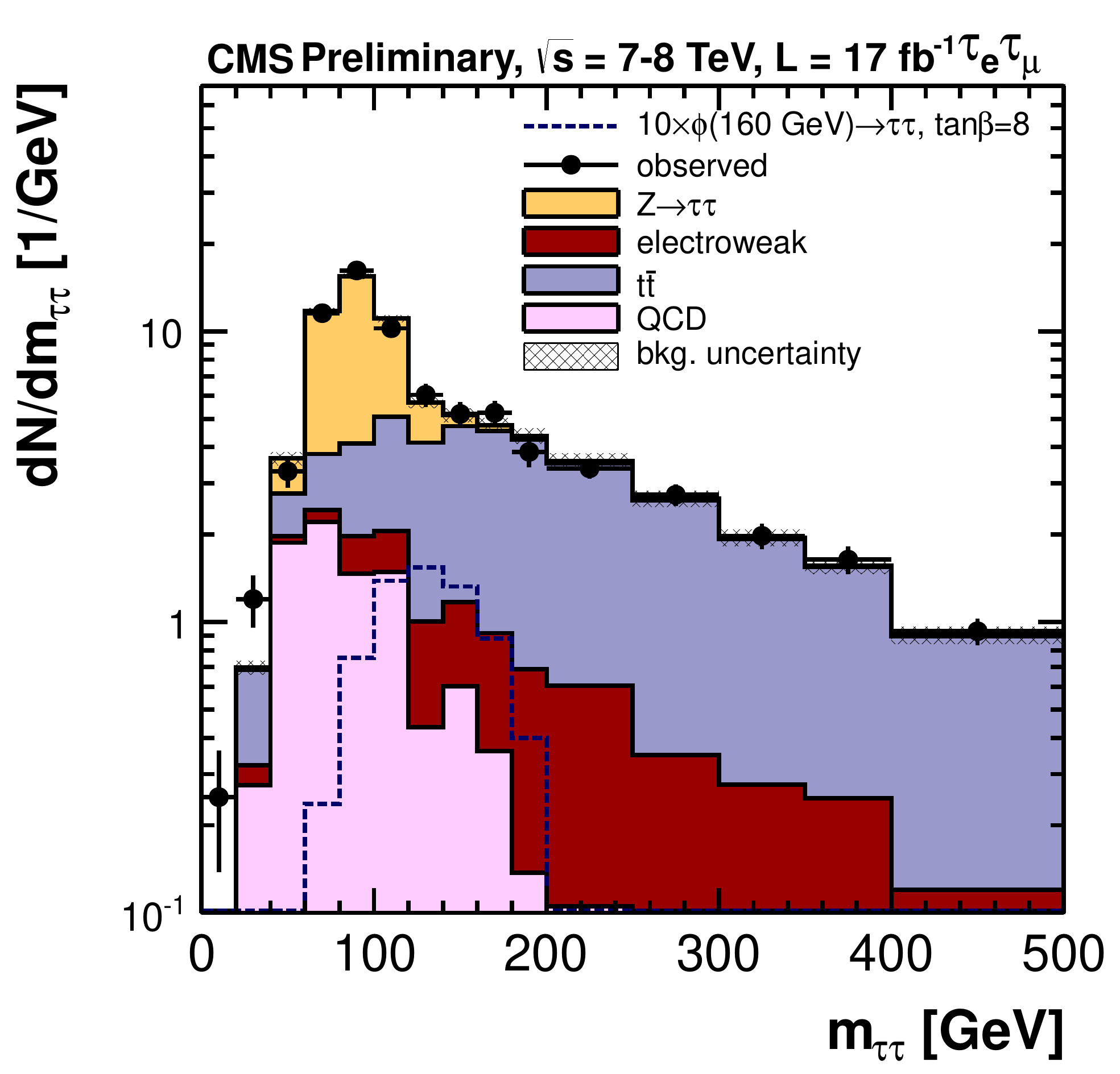}
\includegraphics[width=6.5cm, clip]{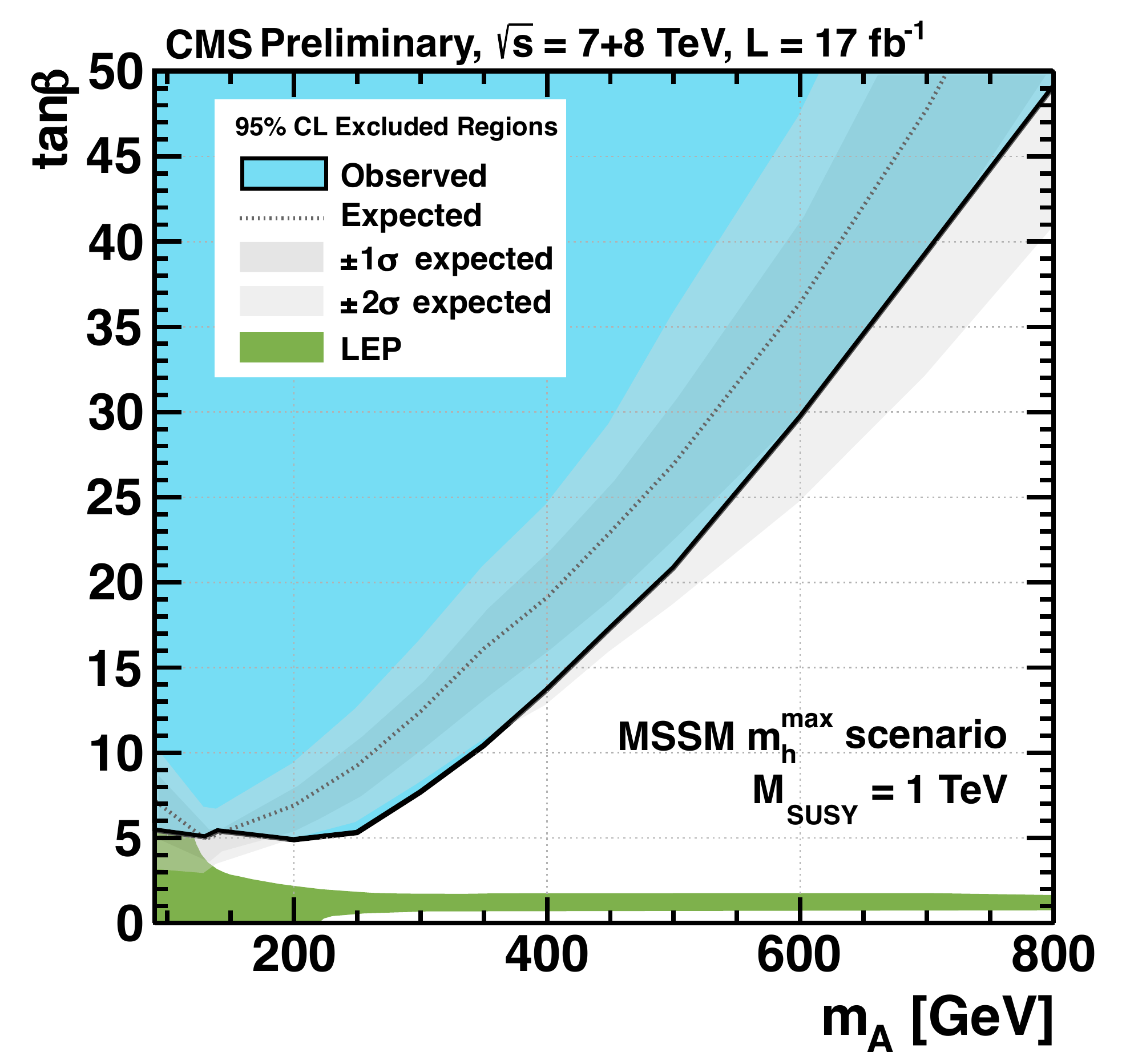}

\caption{Left: The maximum likelihood mass for the $b$-tagged $\tau_\mathrm{e}\tau_\mathrm{\mu}$ selection.
The data are compared to the background expectation and a hypothetical MSSM signal with $m_A = 160$~GeV and $\tan\beta = 8$.
         Right: Expected and observed 95\% confidence level exclusion limits as a function of $m_A$ and $\tan\beta$
for the $m_h^\mathrm{max}$ scenario for the CMS $\Phi \to \tau\tau$ search.~\cite{CMStautau}}
\label{fig:neutral5}       
\end{figure*}

The maximum likelihood mass for the $\tau_\mathrm{e}\tau_\mathrm{\mu}$ $b$-tagged final state is shown in
Figure~\ref{fig:neutral5} (left).  A binned profile likelihood method, using the maximum likelihood mass 
distributions in all subchannels is used to extract 95\% CL exclusion limits, as no significant excess 
of data is observed.  The exclusion limits as a function of $m_A$ and $\tan\beta$ in the $m_h^\mathrm{max}$ 
scenario is shown in Figure~\ref{fig:neutral5} (right).





\section{Summary}
\label{sec:Summary}

Both ATLAS and CMS have carried out searches for MSSM Higgs bosons in high energy $pp$ collision data in a large variety
of search channels.  So far no evidence for Higgs bosons beyond the SM framework exists.  The exclusion limits that have 
been set by both collaborations provide unique and strong constraints on the allowed MSSM parameter space.  
In addition, there are other channels where searches for
MSSM bosons have been made that have not been detailed in these proceedings~\cite{OtherSearch}.
While the MSSM is becoming more and more constrained, future searches using more data will have increased 
sensitivity in the MSSM parameter space, and are highly anticipated by the particle physics community.

%
%
%

\end{document}